\documentstyle[prl,aps]{revtex}

\begin{document}
\title{Disorder--to--order transition in the magnetic and electronic properties of URh$_{2}$Ge$_{2}$}
\author{S. S\"{u}llow$^{1,2,5}$, S.A.M. Mentink$^{3,}$\cite{ment}, T.E. Mason$^{3,}$\cite{mas}, R. Feyerherm$^{4}$, G.J. Nieuwenhuys$^{1}$, A.A. Menovsky$^{1,}$\cite{men}, and J.A. Mydosh$^{1}$}
\address{$^{1}$Kamerlingh Onnes Laboratory, Leiden University, P.O. Box 9506, 2300 RA Leiden, The Netherlands \\
$^{2}$Max--Planck--Institut f\"{u}r Chemische Physik fester Stoffe, Bayreuther Str. 40/Haus 16, 01187 Dresden, Germany \\
$^{3}$Department of Physics, University of Toronto, 60 Saint George Street, Toronto, Ontario, Canada M5S 1A7 \\
$^{4}$Hahn--Meitner--Institut GmbH, Glienicker Str. 100, 14109 Berlin, Germany \\
$^{5}$Institut f\"{u}r Metallphysik und Nukleare Festk\"{o}rperphysik, TU Braunschweig, Mendelssohnstr. 3, 38106 Braunschweig, Germany.}
\date{\today}
\maketitle

\begin{abstract}
We present a study of annealing effects on the physical properties of tetragonal single--crystalline URh$_{2}$Ge$_{2}$. This system, which in as--grown form was recently established as the first metallic 3D random--bond heavy--fermion spin glass, is transformed by an annealing treatment into a long--range antiferromagnetically (AFM) ordered heavy--fermion compound. The transport properties, which in the as--grown material were dominated by the structural disorder, exhibit in the annealed material signs of typical metallic behavior along the crystallographic $a$ axis. From our study URh$_{2}$Ge$_{2}$ emerges as exemplary material highlighting the role and relevance of structural disorder for the properties of strongly correlated electron systems. We discuss the link between the magnetic and electronic behavior and how they are affected by the structural disorder.
\end{abstract}


The magnetic and electronic properties of disordered intermetallic compounds have been the focus of a large number of investigations (for reviews see~\cite{cargill,dugdale,mott,mydosh}). These materials are model systems to study "glassiness", which is observed in the magnetic behavior of spin glasses and the electronic transport of metallic glasses. Surprisingly however, and in spite of the long--standing research efforts, a central topic, the transition from glassy to crystalline behavior, which can be accomplished in such materials, has been widely neglected in these studies. 

At present the consensus is that in order to obtain glassy behavior in an intermetallic compound, a critical value of structural disorder must be exceeded. Hence, tuning the structural disorder provides a tool to investigate the transition from glassy to crystalline behavior. The critical disorder value for the transition from glassy to metallic electronic transport is characterized by the Ioffe--Regel criterion~\cite{ioffe}. It distinguishes between the regime of strongly disordered glassy ($\lambda$\,$\ll$\,$R_{ij}$) compared to weakly disordered metallic ($\lambda$\,$\gg$\,$R_{ij}$) transport in metals ($\lambda$\,=\,elastic mean free path; $R_{ij}$\,=\,atomic nearest--neighbor distance). The magnetic exchange in intermetallics is affected in two ways by disorder. First, the atomic randomness disturbs the spin correlations, leading eventually to a transition from a long--range ordered to a spin--glass state. Secondly, if the disorder is strong enough to cause substantial electronic localization, it suppresses the conduction--electron mediated magnetic exchange.

Both the transitions from glassy metallic and magnetic to crystalline and ordered behavior lead to unusual physical properties. Because of the crystalline disorder the usual Boltzmann--equation based view of electronic transport is no longer appropriate, while the magnetic exchange is randomized and weakened. From a theoretical standpoint, the effect of disorder on transport and magnetic exchange in these limits is only partially understood, while the problem of the interplay between local--moment magnetism and disordered electronic transport is unsolved~\cite{mott,jagannathan,miranda,castro}. Experimentally, owing to a lack of suitable materials these transitions are largely unexplored. In this context we present our case study on the effect of annealing on the properties of URh$_{2}$Ge$_{2}$. For this material we have been able to gradually tune the ground state of the system from a disordered electronic and magnetic into a long--range ordered one by means of annealing. This compound therefore permits to investigate in detail the order--disorder transitions in the magnetic and electronic properties of a local--moment disordered metal, as sketched above.

Previously, we characterized as--grown URh$_{2}$Ge$_{2}$ as a 3D random--bond heavy--fermion spin glass~\cite{sullow}. Based upon x--ray and neutron diffraction studies crystallographic disorder results from a mixing of Rh and Ge atoms over their available lattice sites, while the U atoms are positioned on an ordered $bct$ sublattice. The structural disorder on an atomic scale generates spin--glass behavior via random and competing magnetic interactions. In this contribution we will show that varying the disorder level by metallurgical treatments, like annealing, dramatically affects transport and magnetic properties. In the transport properties a transition from that in a disordered medium to typical metallicity is observed, while magnetically the system is tuned into a long--range ordered AFM state. Our results imply that, while in single--crystalline, as--grown URh$_{2}$Ge$_{2}$ the structural disorder generates the glassy transport and magnetic behavior, the system is close to both a metallic and long--range ordered state. Minute changes of the level of atomic disorder are sufficient to pass the critical disorder limit and to transform the magnetic and electronic glassy state into a crystalline long--range ordered state. To investigate the relationship between magnetic and electronic ground state we therefore performed a thorough study of the physical properties of annealed, single--crystalline URh$_{2}$Ge$_{2}$ and compared it to those of as--grown material. 

The experiments presented here have been carried out on the crystal investigated in Ref.~\cite{sullow}, where details regarding crystal growth and characterization can be found. The composition of the as--grown crystal was determined by electron probe microanalysis (EPMA) to be single phase URh$_{2.00 \pm 0.06}$Ge$_{1.96 \pm 0.06}$. Initially, the physical properties of the crystal were investigated in as--grown form, after which it was annealed, first at 900$^{\circ}$C for 1 week, and subsequently at 1000$^{\circ}$C for a second week. After each heat treatment the main physical and metallurgical properties were determined. No stoichiometry changes occurred with the annealing at 900$^{\circ}$C. However, after the heat treatment at 1000$^{\circ}$C the single crystal was coated with a thin layer ($\sim$\,50\,$\mu$m) of an U rich phase, and small amounts of Rh and Ge were evaporated from the sample. X--ray Laue diffraction proved the sample still to be single--crystalline tetragonal, but with EPMA a small change of the matrix composition to URh$_{1.97 \pm 0.06}$Ge$_{2.08 \pm 0.06}$ was established. To minimize the contributions from the U rich surface phase the crystal was polished to remove as much coating as possible. From x-ray powder diffraction performed on the as--grown and the twice annealed crystal we observe small intensity changes of a number of Bragg peaks caused by the annealing. Unfortunately, the intensity changes are too weak to unambiguously relate them to structural modifications due to the annealing. To resolve the relationship between annealing and structural properties further studies are underway and their results will be presented in due time~\cite{feyerherm}. However, the lack of such structural information does not affect our discussion of the physical properties, as we independently determine the disorder level in the annealed crystals from the physical properties. In the following we refer to the crystal in as--grown form as S1, after annealing at 900$^{\circ}$C as S2 and after annealing at 1000$^{\circ}$C as S3.

We determined the dc-- and ac--susceptibilities $\chi_{dc}$ and $\chi_{ac}$, the former as function of temperature $T$ and field $B$, the latter as function of $T$ and frequency $\omega$, the $T$ dependence of the specific heat $c_{p}$ and the resistivity $\rho$ for the samples S1, S2 and S3. The susceptibilities were obtained in a commercial SQUID, in 0.6\,T between 5 and 300\,K, and at other fields up to 5\,T between 5 and 50\,K. $\chi_{ac}$ was measured in the frequency range 1\,--\,10$^{3}$\,Hz from 5 to 30\,K with a driving field $B_{ac}$\,=\,3$\times$10$^{-4}$\,T. The specific heat was measured using a home--built, semi--adiabatic technique, between 2.5 and 30\,K for the crystals S1 and S2, and between 4 and 25\,K for S3. The resistivity was determined employing a 4--point ac--technique between 1.3 and 300\,K. Because of the annealing--induced phase segregation in crystal S3, the data taken on this crystal could be slightly affected by the stoichiometry change. Still, as our metallurgical and structural analysis of S3 establishes the stoichiometry to be almost 1:2:2 and the crystallographic structure properly tetragonal, important qualitative and semi--quantitative conclusions can be drawn from a comparison of the data on the crystal after the second heat treatment and the as--grown one. 

In Fig.~\ref{fig:fig1}(a) and (b) we plot the dc--susceptibilities as $\chi_{dc}(T)$ and $\chi_{dc}^{-1}(T)$ of S1--S3 for both crystallographic directions in an applied field $B$\,=\,0.6\,T~\cite{alignment}. At all annealing stages a magnetic anisotropy of a factor of 3\,--\,4 between $a$ and $c$ axis is observed, with Curie--Weiss like behavior at high temperatures. Curie--Weiss fits to the data of the three crystals above 100\,K yield values of the Curie--Weiss temperature $\Theta_{CW}$\,/\,effective magnetic moment $\mu_{eff}$ of --120 to --150\,K\,/\,2.9 to 3\,$\mu_{B}$ along the $a$ and --26 to --36\,K\,/\,3.15 to 3.25\,$\mu_{B}$ along the $c$ axis, respectively. It implicates that within the error from alignment variations the high temperature susceptibility does not significantly depend on annealing. Further, since for both crystallographic directions and all annealing stages $\mu_{eff}$ is smaller than expected for a free U$^{3+}$ or U$^{4+}$ ion, it indicates that even at highest temperatures the uranium crystalline electric field (CEF) levels are not equally populated. This accounts for the unphysically large negative values of $\Theta_{CW}$ along the $a$ axes. 

While the single ion properties of the paramagnetic U ions at high temperatures are not affected by the annealing, the nature of the low--temperature magnetic state is transformed from glassy to long--range ordered. This is illustrated in the Figs.~\ref{fig:fig2} and \ref{fig:fig3}, where we display $\chi_{dc}(T)$ of S1--S3 measured in zero--field--cooled (ZFC) and field--cooled (FC) mode and as function of field $B$, and in Fig.~\ref{fig:fig4} depicting $\chi_{ac}(T)$ as function of frequency $\omega$. 

In the FC/ZFC experiment (Fig.~\ref{fig:fig2})~\cite{sullow} (applied field $B$\,=\,5$\times$10$^{-3}$\,T) the as--grown crystal S1 shows for both crystallographic directions the archetypical signs of spin--glass freezing: cusps at 9.3\,K in $\chi_{dc}$ and large irreversibility below the cusps between FC and ZFC experiment, in contrast to the reversible behavior in the paramagnetic phase above the cusps. As for canonical spin glasses like \underline{Cu}{\bf Mn}~\cite{mydosh}, $\chi_{dc}$ below the cusps increases with $T$ for the ZFC run, while it is almost constant in the FC experiment. From the irreversibility point, the temperature at which FC and ZFC run deviate from each other, we determine the freezing temperature $T_{F}$\,=\,9.16\,K.

Annealing the crystal at 900$^{\circ}$C (S2) reduces the tell--tale marks of spin--glass freezing, though they are not completely suppressed. Maxima are visible for both crystallographic directions, but now at 15\,K. Irreversibility is observed between FC and ZFC measurement, though to a much lesser degree than for S1. In addition, the maxima are much broader than in the as--grown case, and $\chi_{dc}$ of the FC measurement has a clear $T$ dependence, which is uncharacteristic for typical spin glasses far into the frozen state. Also unlike typical spin glasses, the irreversibility point lies above the maximum at $T_{irr}$\,=\,18\,K.

The second heat treatment (S3) removes all signs of spin--glass freezing. For $a$ and $c$ axes antiferromagnetic anomalies are visible, with $T_{N}$\,=\,13.3\,K for both directions, determined from the maxima in $d(\chi T)/dT$, and without irreversibility between FC and ZFC experiment. The anomalies are sharp, indicating a well--defined long--range antiferromagnetic transition. Altogether, the FC/ZFC experiments suggest that with the annealing a transition from a spin--glass state in S1 has been achieved towards a long--range antiferromagnetically ordered one in S3. The crystal S2 represents an intermediate state between those two extremes.

This observation is corroborated by our study of the field dependence of $\chi_{dc}$ and the frequency dependence of $\chi_{ac}$. The first we present in Fig.~\ref{fig:fig3} for S1--S3 with the field along $a$ and $c$ axes. For S1 there is a strong suppression and broadening of the freezing transition with increasing magnetic field $B$, resembling the behavior of canonical spin glasses~\cite{mydosh}. These effects are much weaker for S2~\cite{note}, and absent for S3, as expected for a long--range antiferromagnetically ordered system. 

The ac--susceptibilities are shown in Fig.~\ref{fig:fig4}. While for S1 we observe the characteristic spin--glass frequency dependence of $T_{F}$ for $a$ and $c$ axes~\cite{sullow}, it is much weaker for S2. Again, as for the dc--experiment, the magnetic anomaly for S2 is broadened, and a frequency dependence only appears below $T_{ac}$\,$\simeq$\,13\,K, thus much lower than the irreversibility point. In addition, no out--of--phase component is detected for S2, confirming the suppression of the spin--glass freezing. Finally, for S3 a magnetic anomaly is present, but the frequency dependencies have vanished and no out--phase signal is detected, thus confirming the nature of the magnetic state as long--range antiferromagnetically ordered.

The transition from a spin--glass to a magnetically ordered state with annealing is also observed in the specific heat $c_{p}$. In Fig.~\ref{fig:fig5}(a) we plot $c_{p}$/$T$ as function of $T$ for S1--S3. The difference of the absolute $c_{p}$ values at high temperatures for the crystals imply that the $T$ dependence of the CEF levels and/or lattice contributions slightly change with annealing. Without further quantitative information about these changes we can only approximate the background contributions to $c_{p}$, thus allowing us a qualitative discussion of the specific heat, and which is quantitatively exact in the low--$T$ limit, with negligible lattice and CEF contributions.

We performed the correction for crystal S1 as described in Ref.~\cite{sullow}, using the non--magnetic analogous compound UFe$_{2}$Ge$_{2}$ as specific heat background. Below 30\,K this contribution is well represented by a Debye lattice specific heat with $\Theta_{D}$\,=\,203\,K. In order to compare S2 and S3 to S1, we assume Debye lattice backgrounds for the two data sets, but with $\Theta_{D}$ varying slightly to achieve that $c_{p}$ of S1--S3 merge at 25\,K. We obtain values of $\Theta_{D}$\,=\,223\,K for S2 and 216\,K for S3; the approximate background contributions are included in Fig.~\ref{fig:fig5}(a) as solid and broken lines. Subtracting the Debye specific heats from the experimental data yield the corrected specific heat $c_{p,cor}$ at the magnetic anomalies, displayed in Figs.~\ref{fig:fig5}(b) 
and \ref{fig:fig6}(a). 

There are substantial qualitative differences of the $T$ dependence of $c_{p}$ of the crystals S1 and S2, on the one hand, and the twice annealed crystal S3, on the other. For S1 the freezing transition is manifested as broad anomaly with a maximum in $c_{p,cor}$ at 12.8\,K ($\approx$\,1.4\,$T_{F}$), resembling the specific heat effects in canonical spin glasses~\cite{mydosh}. A similar anomaly, but now with a maximum at 14.9\,K, is visible for S2. In contrast, a mean--field like magnetic transition is observed for S3 at $T_{N}$\,=\,13.4\,K, while the low--$T$ specific heat $c_{p,cor}$ of S3 is qualitatively different from that of S1/S2. For S1 and S2 at low temperatures $c_{p}$ is best described by $\gamma T + D T^{\kappa}$, with $\gamma$\,=\,116\,mJ/mole\,K$^{2}$, $D$\,=\,25\,mJ/mole\,K$^{\kappa+1}$ and $\kappa$\,=\,1.80 for S1~\cite{sullow,lowcp}. Instead, for S3 we find below 10\,K the common relation for a heavy--fermion antiferromagnet in the magnetically ordered phase, $c_{p}$\,=\,$\gamma T + \beta T^{3}$, with $\gamma$\,=\,200\,mJ/mole\,K$^{2}$ and $\beta$\,=\,3.74\,$\times$10$^{-4}$\,mJ/mole\,K$^{4}$ (Fig.~\ref{fig:fig6}). Altogether, the specific heat verifies the main result of the susceptibility study: a transition from a spin--glass ground state in as--grown URh$_{2}$Ge$_{2}$ to an antiferromagnetically ordered heavy--fermion state in twice annealed material. Crystal S2 represents an intermediate state, with the higher temperature of the maximum in $c_{p,cor}$, compared to S1, corresponding to a higher temperature of the anomaly in $\chi$, while the resemblance of the low--temperature specific heat to that of crystal S1 proves the absence of any true long--range magnetically ordered phase.

Finally, in Fig.~\ref{fig:fig7}(a) and (b) we present the resistivity, plotted as $\rho /\rho_{300 \rm K}(T)$ for S1, S2 and S3. The absolute values of $\rho$ for crystal S1 at 300\,K are 318\,$\mu \Omega$cm along the $a$ and 450\,$\mu \Omega$cm along the $c$ axis. The salient features of the resistivity of the as--grown crystal S1 are ({\em i.}) unusual large values of $\rho$, ({\em ii.}) a large and temperature dependent anisotropy between $a$ and $c$ axis and ({\em iii.}) negative temperature coefficients for both crystallographic directions up to room temperature. 

We have considered various mechanisms causing the such $\rho$--characteristics. The low--temperature resistivity is much larger than the unitary limit and does not show a logarithmic $T$ dependence, ruling out the Kondo effect (not shown). There is no evidence for a gap or pseudogap, since fits to activated behavior of $\rho (T)$ are poor. This is demonstrated in Fig.~\ref{fig:fig8}(a), where we set out $\ln{\rho}$ vs. $T^{-1}$. Further, the maximum metallic resistivity, either estimated from the Ioffe--Regel criterion~\cite{ioffe,remark} 
\begin{equation}
\rho_{max} = \frac{3 \pi^2 \hbar}{k_{F}^{2} e^{2} a} = 190 \mu \Omega \rm cm,
\label{ioffe}
\end{equation}
or from the Mooij rule~\cite{mooij}, 200$\mu \Omega$cm, is much lower than that observed in the experiments, implying an electronic mean free path smaller then interatomic distances and substantial electronic localization. Consequently, it is necessary to interpret the resistivity as arising from crystallographic disorder~\cite{dugdale,mott,lee,bergmann}, which also accounts for the strong sample dependence of $\rho$. The overall behavior of $\rho$ for our crystal is similar to that reported in Ref.~\cite{dirkmaat}, only the absolute values for our crystal are larger by a factor 1.5--2 for $\rho \| c$. The sample dependent $\rho$ reflects the degree of Rh/Ge stacking disorder, which varies with growth conditions. 

A demonstration for the dominating role of crystallographic disorder on the transport properties and its anisotropy comes from our annealing experiments on URh$_{2}$Ge$_{2}$. In Fig.~\ref{fig:fig7}(a) and (b) we include the normalized resistivities of annealed URh$_{2}$Ge$_{2}$, S2 and S3, along $a$ and $c$ axes. Owing to unfavorable sample shapes of the annealed samples the absolute values of $\rho$ could not be determined with high accuracy. At room temperature the values of $\rho$ for S2 and S3 are the same as those of S1, but within a comparatively large experimental error of $\pm$\,20\,\%. It implies that the resistivity values are still large, and that the structural disorder in the crystals has not been removed completely by the annealing~\cite{resistivity}. 

The most striking result is the different effect of annealing on the transport properties along $a$ and $c$ axes, respectively. The $c$ axis normalized resistivity remains almost unaffected by both annealing procedures. But along the $a$ axis the temperature evolution of $\rho$ changes drastically with annealing. While for S2 there is at least the negative temperature coefficient $d\rho /dT$ up to room temperature, qualitatively resembling the behavior of S1, for S3 above 50\,K $d\rho /dT$ changes from negative to positive, implying a transition from almost insulating to metallic behavior generated by the annealing. In addition, the normalized resistivity exhibits a small anomaly at the antiferromagnetic transition at $T_{N}$\,=\,13.5\,K. 

At present, there is no consensus about the mechanisms causing the unusual transport properties of disordered strongly correlated electron systems~\cite{dugdale,mott,miranda,castro,lee}. For weakly correlated disordered metals the low--$T$ conductivity $\sigma$ to lowest order is predicted to evolve like~\cite{lee}
\begin{equation}
\sigma = \sigma_{0} + a T^{\frac{p}{2}} + b \sqrt{T}.
\label{sigma}
\end{equation}
$p$\,(=\,$\frac{3}{2}$, 2 or 3) depends on the dominant inelastic collision mechanism, the $\sqrt{T}$--term represents corrections from electron--electron interactions to $\sigma$. As illustrated in the double--logarithmic plot of $\sigma - \sigma_{0}$ vs. $T$ for as--grown URh$_{2}$Ge$_{2}$ (Fig.~\ref{fig:fig8}(b)), up to about 20\,K we observe $\sigma - \sigma_{0}$\,$\propto$\,$T^{x}$ with $x$\,$\approx$\,1 for both crystallographic directions. At high $T$ ($>$\,50\,K) for $I$\,$\|$\,$a$ the exponent $x$ changes to $\approx$\,0.5, while for $I$\,$\|$\,$c$ it is closer to 0.7. Similar to URh$_{2}$Ge$_{2}$, different $T^{x}$ regimes of $\sigma$ have been observed for metallic glasses~\cite{mott}. It has been attributed to inelastic electron--electron collisions causing $\sigma$\,$\propto$\,$T$ at low temperatures, while at high $T$ electron--phonon interactions lead to $\sigma$\,$\propto$\,$T^{0.5}$. Yet, this interpretation has been questioned as an oversimplification~\cite{lee}. Moreover, in URh$_{2}$Ge$_{2}$ the situation is more complicated, as we expect an additional magnetic scattering contribution. Further, it is not evident that there is a quantitative one--to--one correspondence between the behavior of a weakly and strongly correlated disordered metal. Therefore, we will limit ourselves here to a phenomenological and qualitative discussion of the resistivity, while an extensive discussion of the resistivity we will present elsewhere~\cite{res}.

The annealing procedure does not fundamentally modify the band structure or related properties of URh$_{2}$Ge$_{2}$. If the annealing would change these properties, it should equally affect the resistivities along the $a$ and the $c$ axis. In contrast, we observe a $c$ axis transport independent of annealing, together with an $a$ axis resistivity changing from an almost insulating to a metallic behavior. Consequently, the resistive properties of as--grown and annealed URh$_{2}$Ge$_{2}$ are mainly caused by the type and degree of structural disorder, and not by the underlying band structure of a "perfectly well ordered" URh$_{2}$Ge$_{2}$. In particular, the metallic resistivity along the $a$ axis in twice annealed URh$_{2}$Ge$_{2}$ indicates that the disorder is more strongly reduced within the tetragonal plane than along the $c$ axis. Since the susceptibility proves that annealing does not affect the ionic properties of the U atoms in URh$_{2}$Ge$_{2}$, the crystals after the different heat treatments constitute to good approximation URh$_{2}$Ge$_{2}$ containing three different levels of Rh/Ge disorder. This implies that our compound allowed us to study the problem set out in the introduction, {\em viz.} the transition from glassy to crystalline electronic and magnetic behavior as the degree of structural disorder is varied. 

In as--grown form, because of the atomic scale disorder, the system behaves glassy with respect to the electronic transport and magnetic properties. Annealing the crystal reduces the disorder. For moderate annealing, {\em i.e.} for S2, the glassy behavior still dominates, although regarding its magnetic properties the system cannot be properly described as a pure spin glass anymore, but rather as a mixture of AFM clusters and spin glass. This is evidenced by the broad magnetic anomaly and the large difference between the irreversibility temperature $T_{irr}$\,=\,18\,K from the FC/ZFC experiment and the temperature $T_{ac}$\,$\approx$\,13\,K, below which a frequency dependence of $\chi_{ac}$ is observed. This suggests that the system consists of magnetically correlated regions with a wide distribution of sizes, ranging from single spins leading to the spin glass frequency dependence of $\chi_{ac}$ to large magnetic clusters, blocking below 18\,K and causing the difference between FC and ZFC data at these temperatures. That the freezing/blocking temperatures increase for S2 compared to S1 is simply related to the reduced disorder in S2. 

While the average size of the magnetically correlated regions is larger in S2 than in S1, there are no long--range ordered regions in the crystal. Hence, the specific heat still exhibits the broad anomaly that is characteristic for short--range magnetic phenomena, and the temperature dependence of $c_{p}$ does not follow a $T^{3}$ behavior, as it would have been expected for antiferromagnetic magnons. However, the $T$ dependence of the specific heat of S1 and S2 is also not in agreement with the prediction of the two--level--model, $c_{p}$\,$\propto$\,$T$~\cite{mydosh,anderson}. The experimentally observed intermediate exponent $c_{p}$\,$\propto$\,$T^{1.9}$, together with the similarity of the specific heats of S1 and S2 might therefore be taken as qualitative argument for dimensionally reduced magnons within the magnetic clusters causing such $T$ dependence of $c_{p}$. 

For a sufficiently long annealing treatment, that is for S3, the disorder is reduced to a degree that a long--range ordered magnetic, and along the $a$ axis, a crystalline metallic state is realized. We note that in recent neutron diffraction experiments performed on S3 the antiferromagnetic long--range ordered structure has been directly observed, with an ordered moment of 0.5\,$\mu_{B}$ pointing along the c--axis; details of these investigations will be published elsewhere~\cite{feyerherm}. The fact that along the $c$ axis the resistivity still exhibits the characteristics of transport in a disordered medium suggests that with the annealing a state is created, in which the localization of the conduction electrons in the $c$ direction is much stronger than within the tetragonal plane. 

The observation of a magnetic anomaly in $\rho/\rho_{300\rm K}$ along the $a$ axis in S3 at $T_{N}$, in contrast to the absence of such an anomaly along the $c$ axis for the same crystalline piece, is surprising and warrants further exploration. It suggests that critical fluctuations depend on the electronic mean free path or the extent of the electronic wave functions. In our case, as long as the mean free path or the electron wave functions are smaller than the magnetic lattice spacing (which represents a lower cut--off length scale for the fluctuations), no critical magnetic fluctuations are observable. For twice annealed URh$_{2}$Ge$_{2}$ this is the case for the resistivity along the $c$ axis. If however the mean free path or wave functions extend over a few magnetic lattice sites, as for the resistivity along the $a$ axis, critical fluctuations can be observed. Theoretically, to our knowledge this feature has never been investigated, and we hope that our result initiates efforts to solve this problem.

With respect to magnetism, we must account for the complete replacement of spin--glass by long--range order. To create a spin--glass state in a material with RKKY magnetic exchange, as it is generally the case for $f$--electron intermetallics, the total RKKY--amplitude from the competing interactions must average to near zero. This balance of ferro-- and antiferromagnetic exchange is present in as--grown URh$_{2}$Ge$_{2}$. The replacement of the spin--glass ground state by long--range magnetic ordering in annealed URh$_{2}$Ge$_{2}$ indicates a shift of the balance. We have no experimental access to measure the local magnetic exchange, but we can imagine a simple mechanism causing such effects. 

Our mechanism is based on the assumption, that with annealing the length scale, over which the magnetic exchange is effective, is increased. From the resistivity we know that the magnetic interaction is primarily modified by the annealing in the tetragonal plane, and therefore we consider the two dimensional problem of the in--plane interaction. In the as--grown spin--glass material we have a balance of ferro-- and antiferromagnetic exchange within the plane. Let us assume that this is realized by a ferromagnetic interaction $J_{FM}$ along the unit--cell axes with the nearest neighbors, an antiferromagnetic exchange $J_{AFM}$ along the unit--cell diagonal with the next--nearest neighbors, and with the condition \,$J_{FM}$\,=\,$J_{AFM}$ to ensure balanced competing interactions~\cite{model}. If now the magnetic interaction length scale is increased with the annealing, it implies that additional magnetic interactions from next--next--nearest neighbors etc. have to be taken into account. Then, it is obvious that even if we retain the condition $J_{FM}$\,=\,$J_{AFM}$ in the annealed material, the additional magnetic interactions can shift the balance and in effect create a long--range ordered magnetic state.

So far, theoretical investigations only treat the RKKY--interaction in the presence of weak disorder~\cite{jagannathan}, which requires that the mean free path $\lambda$ is much larger than interatomic distances. In URh$_{2}$Ge$_{2}$, in contrast, we have strong disorder, with the mean free path of the order of interatomic distances. Unfortunately, in this limit of strong disorder there is no knowledge on the dependence of the magnetic exchange on the mean free path. Qualitatively, at least, it is obvious that there must be a transition region from the disorder independent magnetic exchange for the case of weak disorder to that of fully localized electrons in a strongly disordered medium, which leads to a breakdown of conduction--electron mediated magnetic exchange. We suggest that URh$_{2}$Ge$_{2}$ lies right in this transition region, and that with annealing we tune the length scale of the effective magnetic interaction. Again, we hope that our experiments motivate theoretical efforts on the magnetic exchange in the strong disorder limit, even though we are aware that this is an extraordinary difficult task.

Our scenario suggests that spin--glass behavior in dense magnetic compounds like URh$_{2}$Ge$_{2}$ is by far more likely for systems with only nearest neighbor interactions, which in turn qualitatively implies small electronic mean free paths and small magnetic moments. Indeed, the various reported cases of intermetallic random--bond spin glasses, like U$_{2}$PdSi$_{3}$~\cite{resistivity} or PrAu$_{2}$Si$_{2}$~\cite{hemberger}, are systems with large resistivities and without large--moment elements. Also, for PrAu$_{2}$Si$_{2}$~\cite{hemberger} the isoelectronic replacement of Si by Ge has been shown to suppress the spin--glass state and induce antiferromagnetic long--range order, which can be interpreted as arising from the inequivalence of antiferro-- and ferromagnetic ordering upon alloying. But evidently, more experimental studies will be necessary to test the general validity of such a prediction. 

In conclusion, URh$_{2}$Ge$_{2}$ emerges from our study as a model compound that allows us to investigate the role of disorder for the magnetic and electronic properties in strongly correlated electron systems. Our work clearly establishes that disorder plays an important role in such compounds, and that the material responds sensitively on the variation of the disorder level. Further, our study explains the large sample dependencies for URh$_{2}$Ge$_{2}$. Several groups recently observed maxima in the susceptibility, which in retrospective have to be attributed to the spin--glass freezing based upon compound disorder~\cite{sullow,dirkmaat,bak,lloret}. In two reports~\cite{lloret,thompson} even long--range magnetic ordering was reported. All in all, pronounced sample dependencies are observed for almost any physical property. Of course, if annealing at temperatures of the order of 1000$^{\circ}$C is sufficient to transform the spin--glass system into a long--range ordered one, differences in the sample preparation, annealing procedures etc. will specifically affect the physical properties. 

Further, our work outlines future routes of investigations. For instance, it will be interesting to relate the physical properties directly to the structural behavior. As pointed out, a possible scenario for the replacement of the spin--glass state by antiferromagnetic order would be that the anisotropic change of the magnetic exchange strength with annealing destroys the balance between the competing magnetic interactions. Here, a detailed study of the disorder employing microscopic (NMR, M\"{o}\ss bauer--spectroscopy, EXAFS) as integral (neutron and x--ray diffraction) techniques, combined with a determination of the physical properties, would allow a test of such a scenario. Another point of interest is the process of transforming the spin--glass state into the antiferromagnetism. The question, which we cannot answer on basis of our data, is if the transition is continuous, with magnetic clusters growing gradually as the disorder is reduced, or if above a certain cluster size a percolative, long--range ordered state suddenly appears across the crystal. A detailed, and particularly microscopic study of the magnetic state should give new insight into the problem of the glassy and ordered state as either competing or collaborative effects. Finally, a principal problem of disordered magnets is the existence of the (classical) Griffiths phase, a region between the ordering temperatures of the disordered and ordered system~\cite{castro,griffiths}. The possibility to tune the disorder level in URh$_{2}$Ge$_{2}$ should allow to investigate these questions in a much more efficient manner than was previously possible. 

We like to thank B. Becker for performing the specific heat measurement on the twice annealed crystal S3, and F. Galli for the neutron scattering results prior to publication. This work was supported by the Nederlandse Stichting FOM, the CIAR and NSERC of Canada, and the Deutsche Forschungsgemeinschaft DFG. The crystal was prepared at FOM--ALMOS.

\begin{figure}
\caption[1]{Temperature dependence of (a.) the dc--susceptibility $\chi_{dc}$ and (b.) inverse susceptibility $\chi_{dc}^{-1}$ of as--grown (S1; $\Box$) and annealed (S2; $\bullet$ and S3; $\triangle$) URh$_{2}$Ge$_{2}$ in a field of $B$\,$=$\,0.6\,T, applied along $a$ and $c$ axes.}
\label{fig:fig1}
\end{figure}

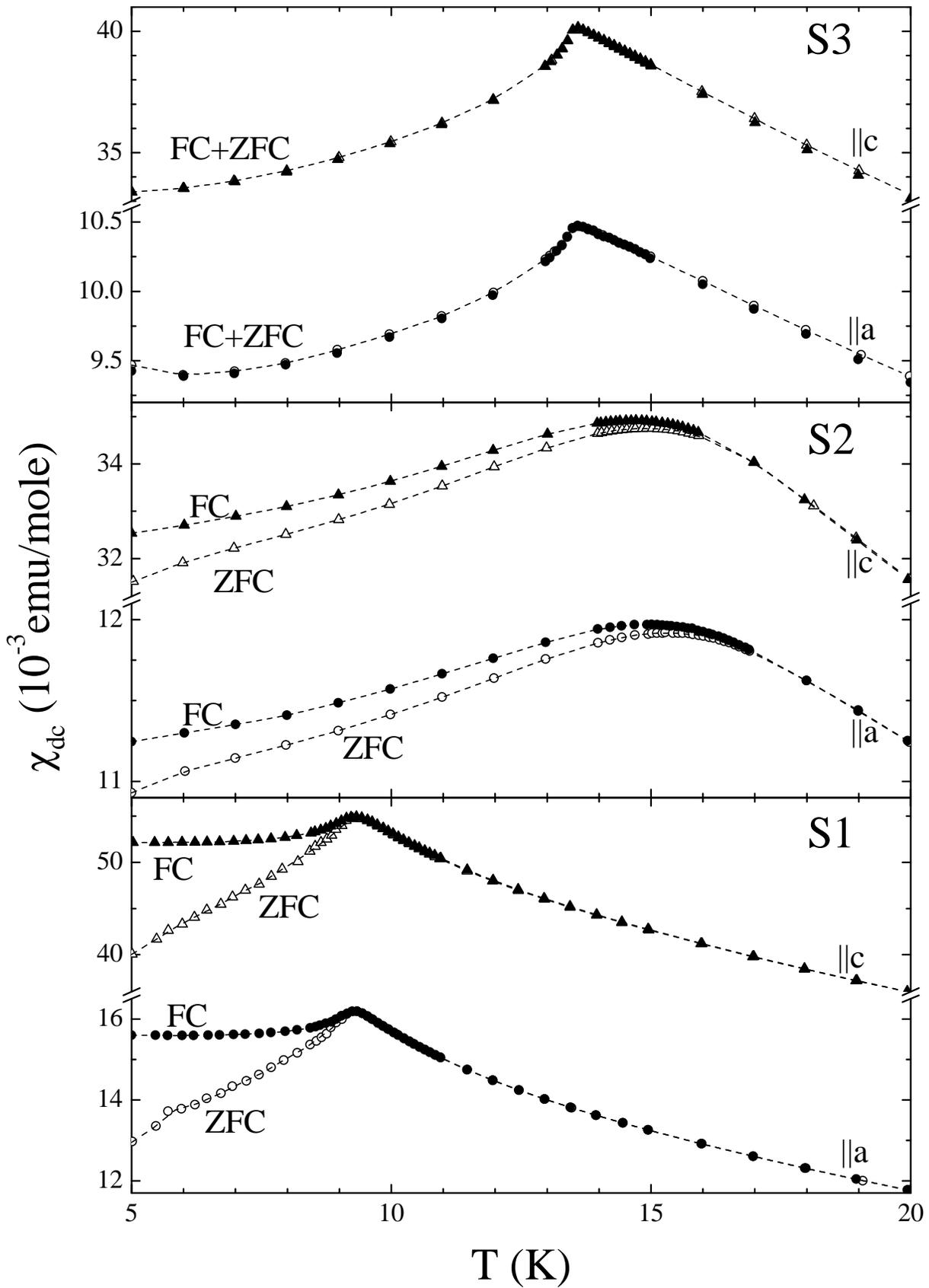
\begin{figure}
\caption[2]{The low--temperature dc--susceptibility $\chi_{dc}$, measured in field--cooled (FC; filled symbols) and zero--field--cooled (ZFC; open symbols) mode for URh$_{2}$Ge$_{2}$ along $a$\,($\circ$) and $c$\,($\triangle$) axis after various heat treatments.}
\label{fig:fig2}
\end{figure}

\begin{figure}
\caption[3]{The field dependence of the dc--susceptibility $\chi_{dc}$ of URh$_{2}$Ge$_{2}$ for both crystallographic directions for (a.) S1, (b.) S2 and (c.) S3 in applied fields of 0.005\,T\,($\Box$), 0.6\,T\,($\circ$), 2\,T\,($\triangle$) and 5\,T\,(+).}
\label{fig:fig3}
\end{figure}

\begin{figure}
\caption[sus4]{The temperature and frequency dependence of the ac--susceptibility $\chi_{ac}$ of URh$_{2}$Ge$_{2}$ for both crystallographic directions for (a.) S1, (b.) S2 and (c.) S3 in $B_{ac}$\,=\,0.3\,mT and at frequencies $\omega$ of 1.157\,Hz\,(solid line), 11.57\,Hz\,($\triangle$), 115.7\,Hz\,($\circ$) and 1157\,Hz\,(+).}
\label{fig:fig4}
\end{figure}
 
\begin{figure}
\caption[5]{(a) The temperature dependence of the specific heat $c_{p}/T$ of URh$_{2}$Ge$_{2}$ for the crystals S1\,($\Box$), S2\,($\triangle$) and S3\,($\circ$). The lines indicate background corrections for S1\,(solid line), S2\,(dotted line) and S3\,(dashed line). (b) The magnetic specific heat $c_{p,cor}$ vs. $T$ of URh$_{2}$Ge$_{2}$ for S1\,($\Box$), S2\,($\triangle$) and S3\,($\circ$); for details see text.}
\label{fig:fig5}
\end{figure}

\begin{figure}
\caption[6]{(a) The temperature dependence of specific heat $c_{p,cor}/T$ of URh$_{2}$Ge$_{2}$ for crystals S1\,($\Box$), S2\,($\triangle$) and S3\,($\circ$). (b) The magnetic specific heat $c_{p,cor}/T$ vs. $T^{2}$ of URh$_{2}$Ge$_{2}$ for S1\,($\Box$), S2\,($\triangle$) and S3\,($\circ$). The lines indicate low--temperature fits to $c_{p}$, for details see text.}
\label{fig:fig6}
\end{figure}

\begin{figure}
\caption[7]{The temperature dependence of the normalized resistivity $\rho /\rho_{300 \rm K}$ of URh$_{2}$Ge$_{2}$ for the crystals S1\,(solid line), S2\,($\circ$) and S3\,($\triangle$) for (a) $I$\,$\|$\,$a$ and (b) $I$\,$\|$\,$c$ axis.}
\label{fig:fig7}
\end{figure}

\begin{figure}
\caption[8]{The resistivity $\rho$ of as--grown URh$_{2}$Ge$_{2}$ (S1) set out in an activation plot $\ln{\rho}$ vs. $T^{-1}$ (a) and as conductivity $\sigma - \sigma_{0}$ vs. $T$ in a double--logarithmic representation (b) along $a$\,($\circ$) and $c$\,($\Box$) axis. The dashed lines denote dependencies $\sigma - \sigma_{0}$\,$\propto$\,$T$ and $T^{0.5}$, respectively.}
\label{fig:fig8}

\end{figure}


\begin{references}
\bibitem[\dagger]{ment} Present address: Philips Research Laboratories, Prof. Holstlaan 4, 5656 AA Eindhoven, The Netherlands.
\bibitem[+]{mas} Present address: Oak Ridge National Laboratory, Oak Ridge, TN 37831, U.S.A..
\bibitem[*]{men} Also at Van der Waals-Zeeman Laboratory, University of Amsterdam, Valckenierstraat 65, 1018 XE Amsterdam, The Netherlands.
\bibitem{cargill} R.J. Cargill III, {\it Solid State Phys.}, Vol. 30, edited by F. Seitz and D. Turnbull, (Academic Press, New York, 1975).
\bibitem{dugdale} J. Dugdale, Contemp. Phys. {\bf 28}, 547 (1987), {\it The electrical properties of disordered metals,} (Cambridge University Press, Cambridge, 1995).
\bibitem{mott} N. Mott, {\it Metal--Insulator Transitions}, (Taylor\&Francis, London, 1990), 2nd ed.. 
\bibitem{mydosh} J.A. Mydosh, {\it Spin Glasses: An experimental Introduction} (Taylor \& Francis, London, Washington, 1993).
\bibitem{ioffe} A.F. Ioffe and A.R. Regel, Prog. Semicond. {\bf 4}, 237 (1960).
\bibitem{jagannathan} A. Jagannathan, E. Abrahams, and M. Stephen, Phys. Rev. B {\bf 37}, 436 (1988); I.V. Lerner, {\it ibid.} {\bf 48}, 9462 (1993).
\bibitem{miranda} E. Miranda, V. Dobrosavljevi\'{c}, and G. Kotliar, Phys. Rev. Lett. {\bf 78}, 290 (1997).
\bibitem{castro} A.H. Castro--Neto, G. Castilla, and B.A. Jones, Phys. Rev. Lett. {\bf 81}, 3531 (1998).
\bibitem{sullow} S. S\"{u}llow {\it et al.}, Phys. Rev. Lett. {\bf 78}, 384 (1997); Physica B {\bf 230--232}, 105 (1997).
\bibitem{feyerherm} R. Feyerherm {\it et al.}, in preparation.
\bibitem{alignment} Small differences of the absolute $\chi_{dc}$ values for the three crystals most likely arise from small variations of the alignment between crystal axes and field direction, which can vary by about $\pm$\,5\,$^{\circ}$.
\bibitem{note} $\chi_{dc}$ for in 0.6\,T $\| a$ has accidentally been taken after cooling the sample in a field of 5\,T, which explains its low--$T$ up--turn.
\bibitem{lowcp} For S2 we find below 7\,K $\gamma$\,=\,108\,mJ/mole\,K$^{2}$, $D$\,=\,9.2\,mJ/mole\,K$^{\kappa+1}$ and $\kappa$\,=\,2.21.
\bibitem{remark} We assume as conduction electron numbers per atom: U\,--\,3; Rh\,--\,1; Ge\,--\,2, implying that $k_{F}$\,=\,1.47\,\AA $^{-1}$; interatomic distances are $a$\,$\approx$\,3\,\AA .
\bibitem{mooij} J.H. Mooij, Phys. Status Solidi A {\bf 17}, 521 (1973).
\bibitem{lee} P.A. Lee and T.V. Ramakrishnan, Rev. Mod. Phys. {\bf 57}, 287 (1985).
\bibitem{bergmann} G. Bergmann, Phys. Rep. {\bf 107}, 1 (1984); C.C. Tsuei, Phys. Rev. Lett. {\bf 57}, 1943 (1986); G. Holter, H. Adrian, and B. Hensel, J. Magn. Magn. Mat. {\bf 63\&64}, 475 (1987); H. Nakamura {\it et al.}, J. Phys.: Cond. Matter {\bf 9}, 4701 (1997); J.M. Barandiar\'{a}n {\it et al.}, {\it ibid.} {\bf 9}, 5671 (1997).
\bibitem{dirkmaat} A.J. Dirkmaat {\it et al.}, Europhys. Lett. {\bf 11}, 275 (1990).
\bibitem{resistivity} Similar large, weakly $T$ dependent resistivities with a negative $d \rho/d T$ are frequently observed in U intermetallics, like in UCu$_{5-x}$Pd$_{x}$ (B. Andraka and G.R. Stewart, Phys. Rev. B {\bf 47}, 3208 (1993)) or U$_{2}$PdSi$_{3}$ (D.X. Li {\it et al.}, Phys. Rev. B {\bf 57}, 7434 (1998)). While in these system crystallographic disorder certainly plays a role, the extent to which disorder controls the transport properties is still a matter of debate. 
\bibitem{res} S. S\"{u}llow {\it et al.}, in preparation.
\bibitem{anderson} P.W. Anderson, B.I. Halperin, and C.M Varma, Philosophical Magazine {\bf 25}, 1 (1972).
\bibitem{model} These assumptions serve as specific example to illustrate the problem. They do not limit the general applicability of our model. We note however that qualitatively the competition of ferro- and antiferromagnetic exchange has been observed in the closely related compound URu$_{2}$Si$_{2}$; see C. Broholm {\it et al.}, Phys. Rev. B {\bf 43}, 12809 (1991). 
\bibitem{hemberger} A. Krimmel {\it et al.}, Phys. Rev. B {\bf 59}, R6604 (1999); J. Phys.: Condens. Matter {\bf 11} (1999) 6991.
\bibitem{bak} H. Ptasiewicz--Bak, J. Leciejewicz, and A. Zygmunt, J. Phys. F: Metal Phys. {\bf 11}, 1225 (1981); Sol. State Commun. {\bf 55}, 601 (1985).
\bibitem{lloret} B. Lloret {\it et al.}, J. Magn. Magn. Mat. {\bf 67}, 232 (1987).
\bibitem{thompson} J.D. Thompson, Z. Fisk, and L.C. Gupta, Phys. Lett. A {\bf 110}, 470 (1985).
\bibitem{griffiths} R.B. Griffiths, Phys. Rev. Lett. {\bf 23}, 17 (1969); A. Brooks Harris, Phys. Rev. B {\bf 12}, 203 (1975); H. Rieger and A.P. Young, Phys. Rev. B {\bf 54}, 3328 (1996); M. Guo, R.N. Bhatt, and D.A. Huse, Phys. Rev. B {\bf 54}, 3336 (1996).
\end{references}
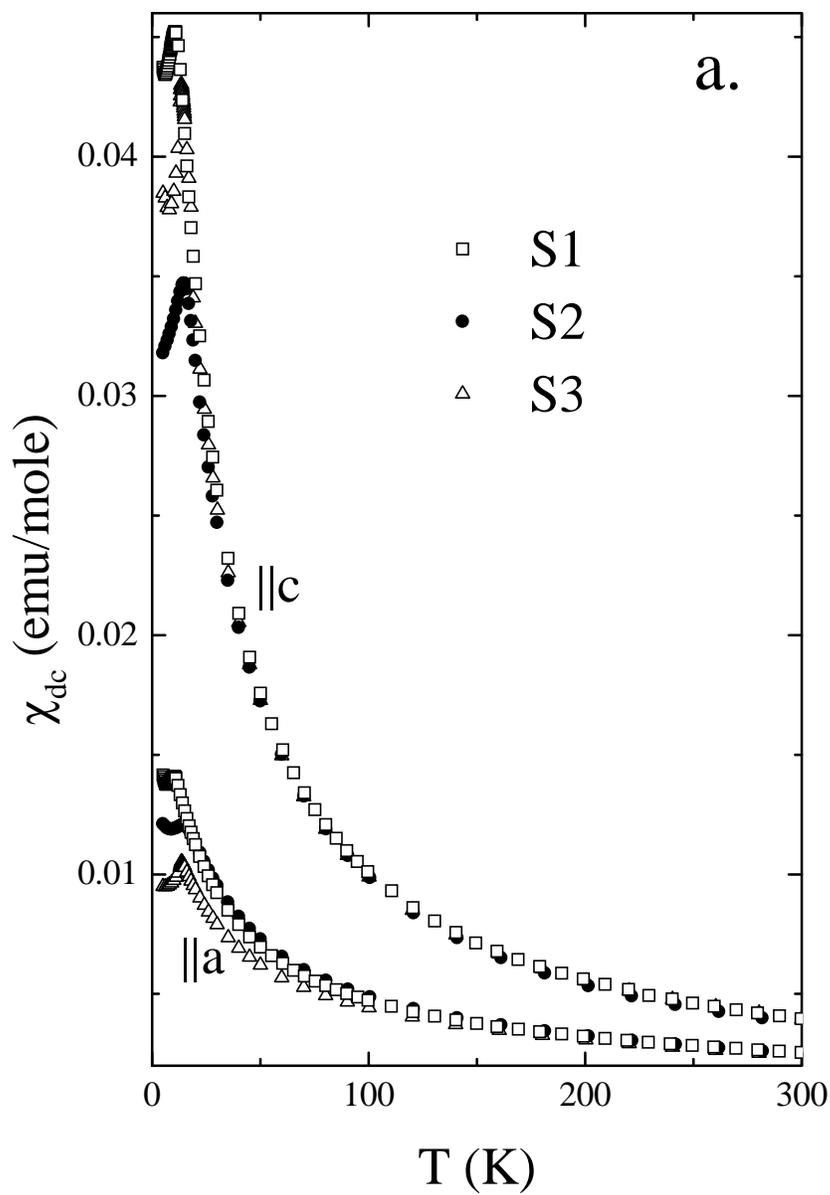
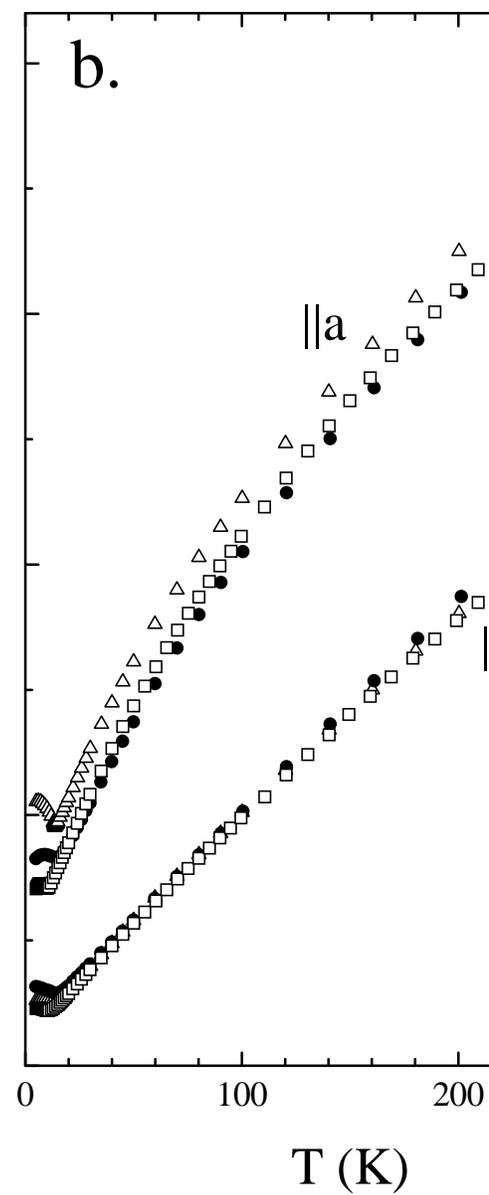
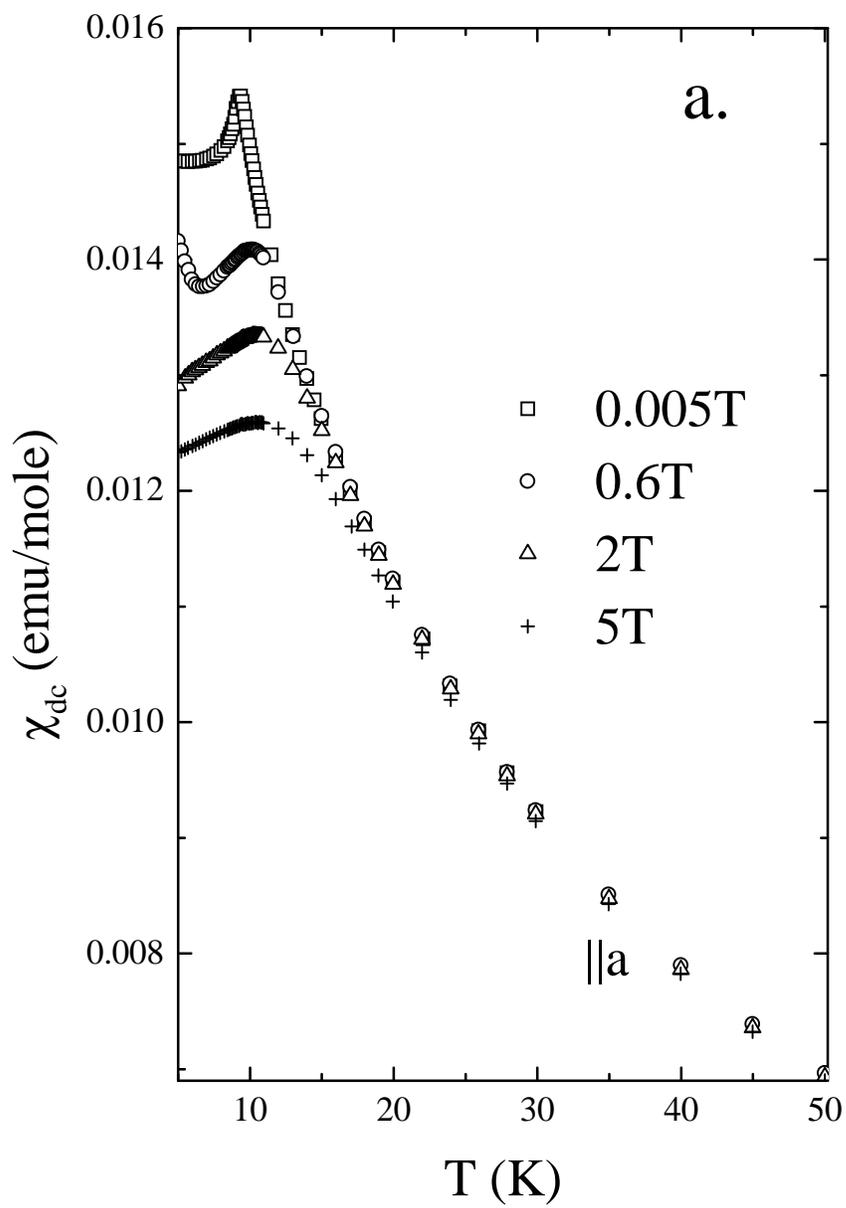
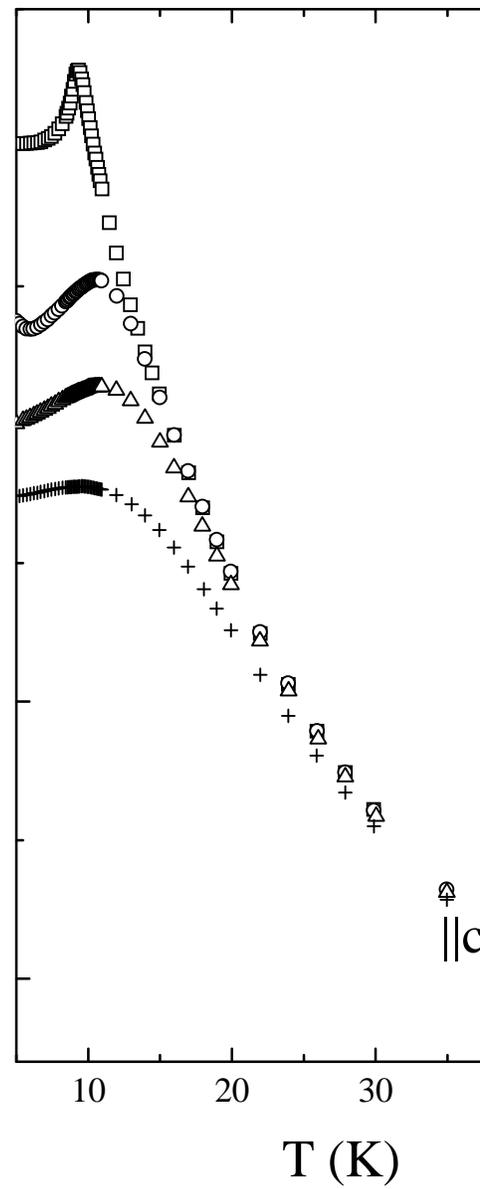
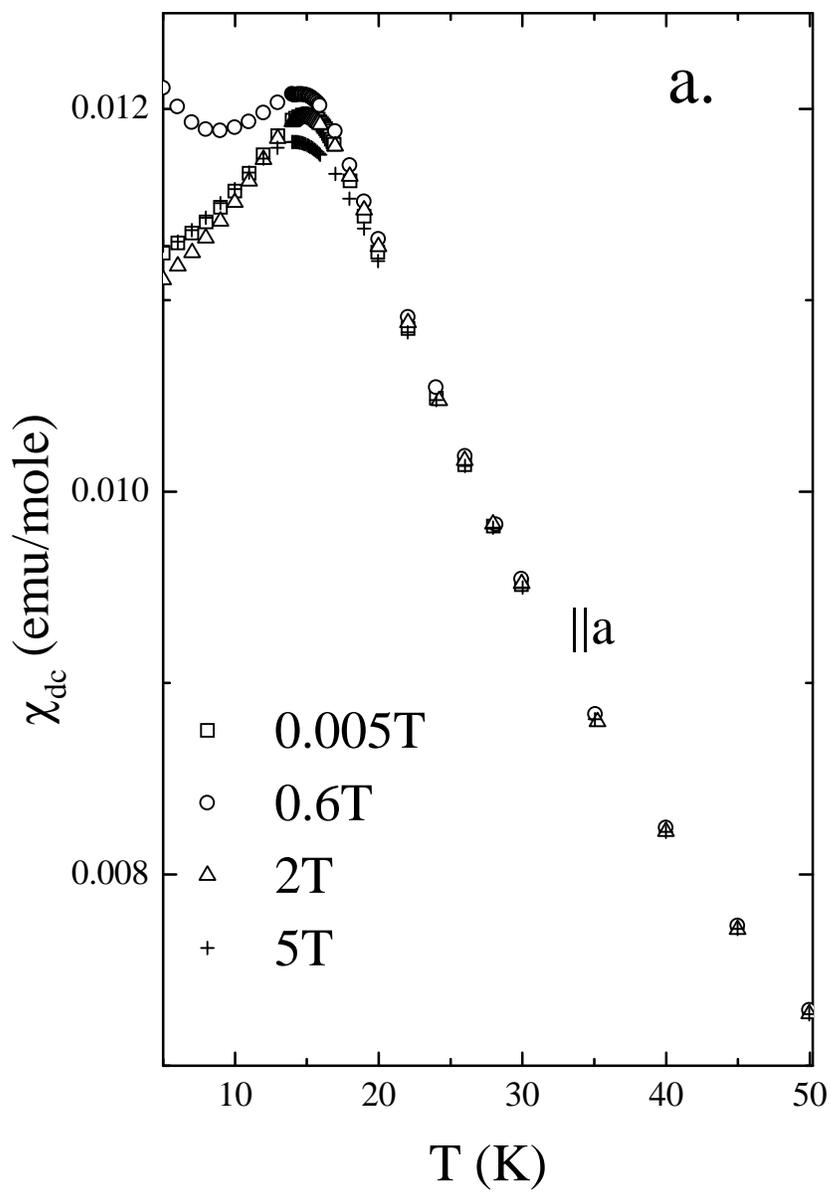
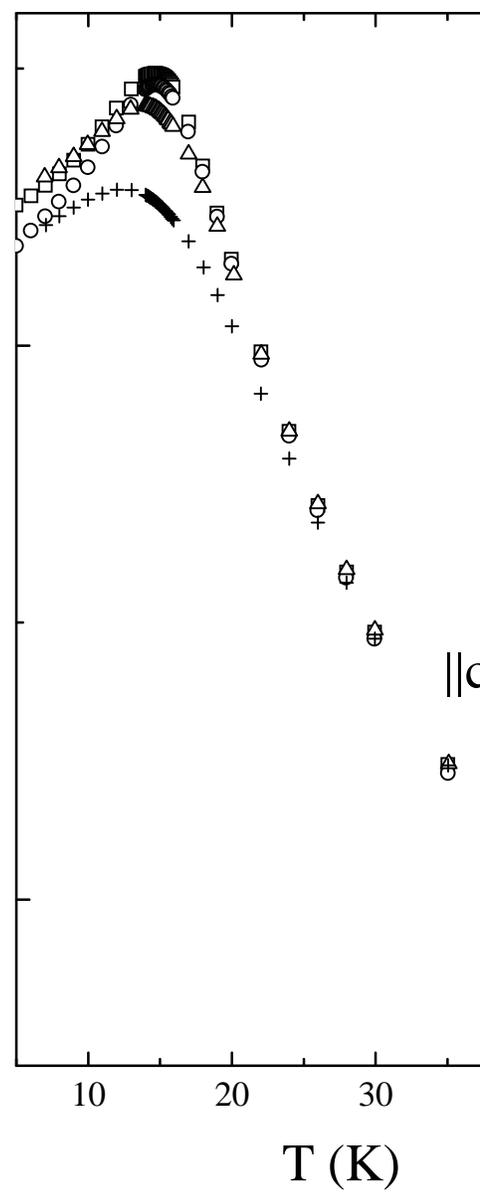
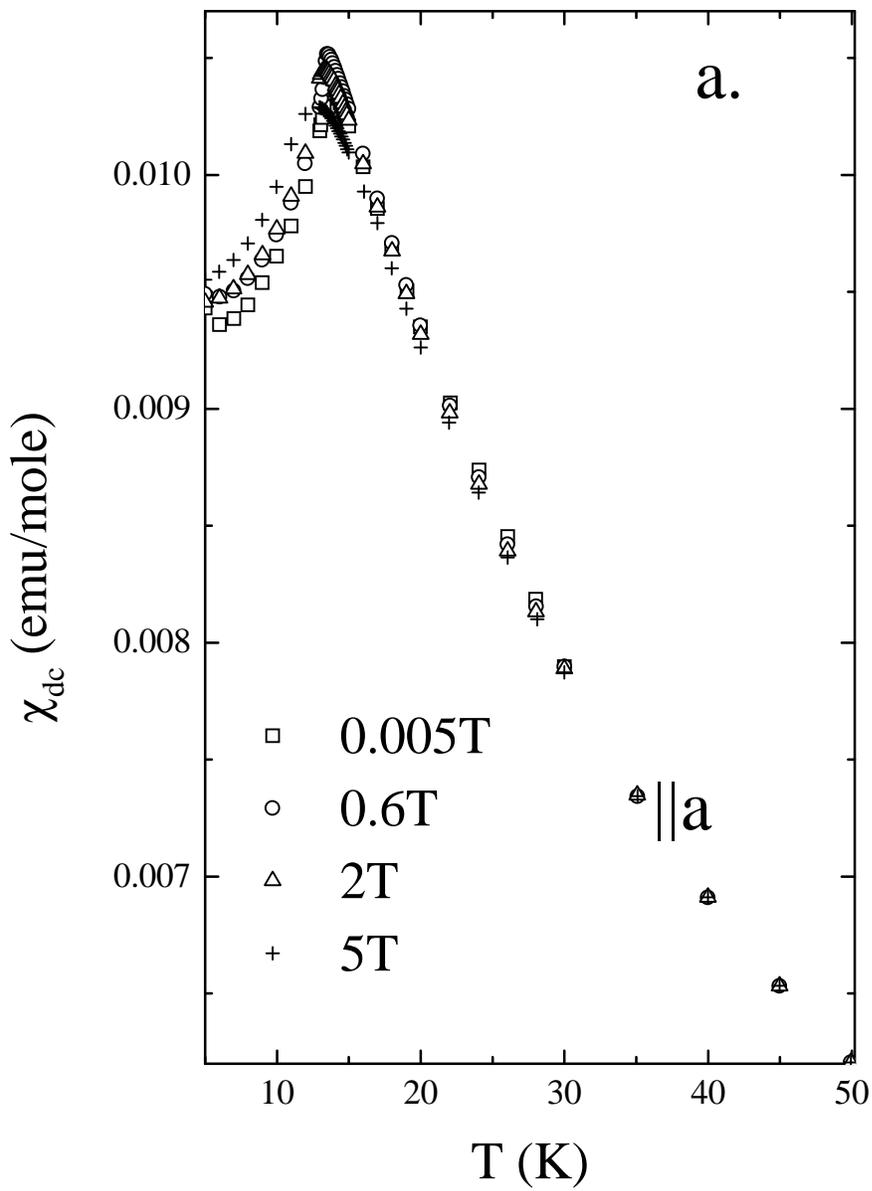
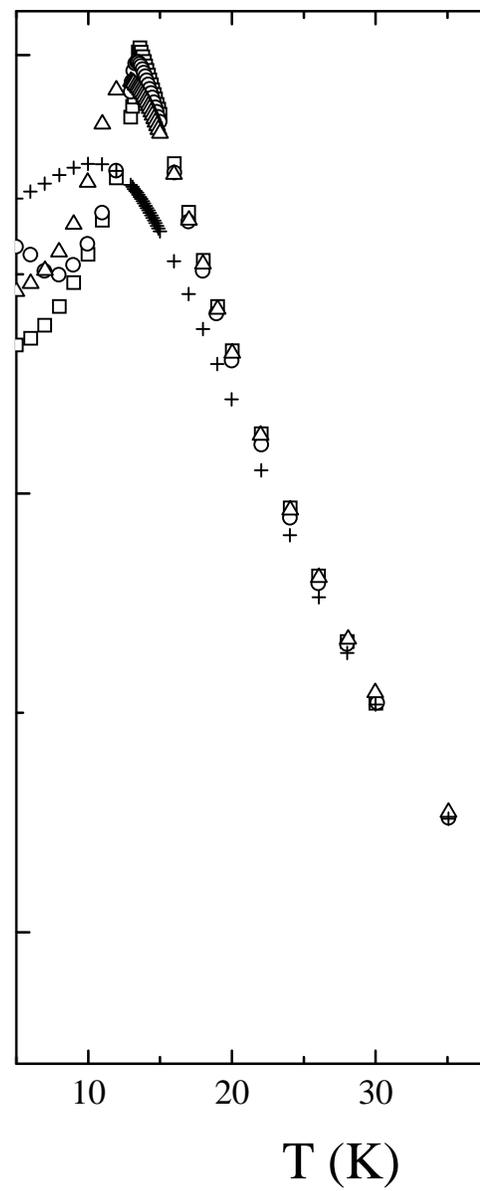
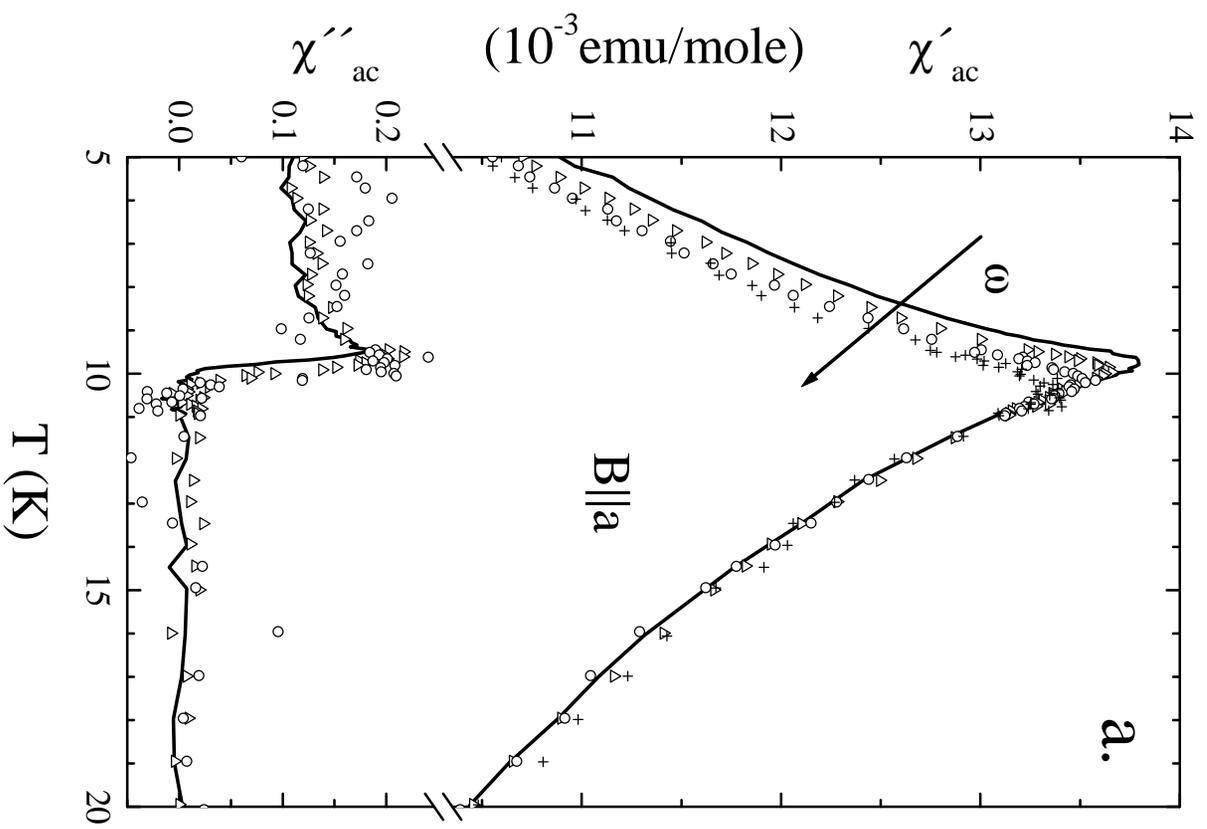
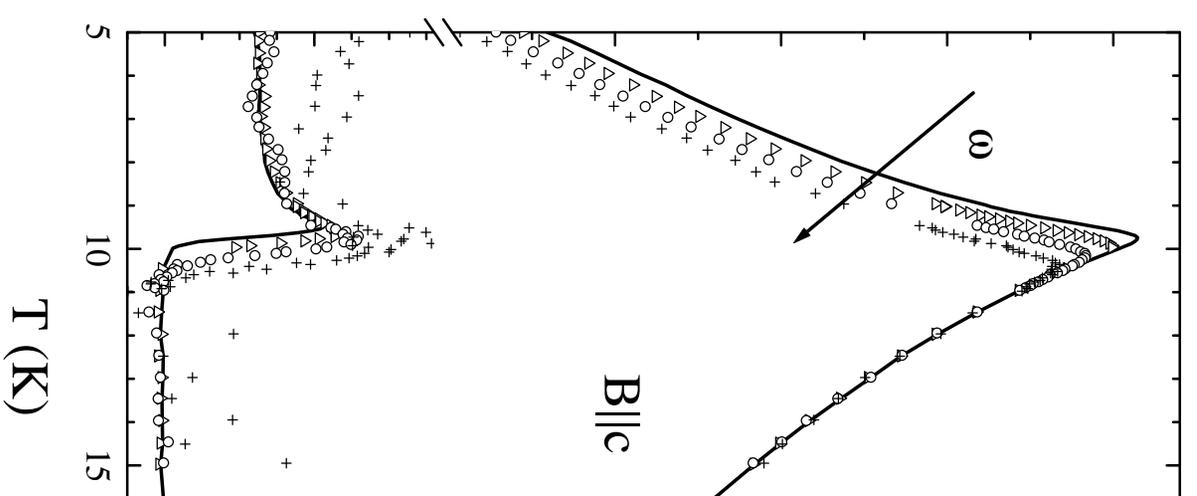
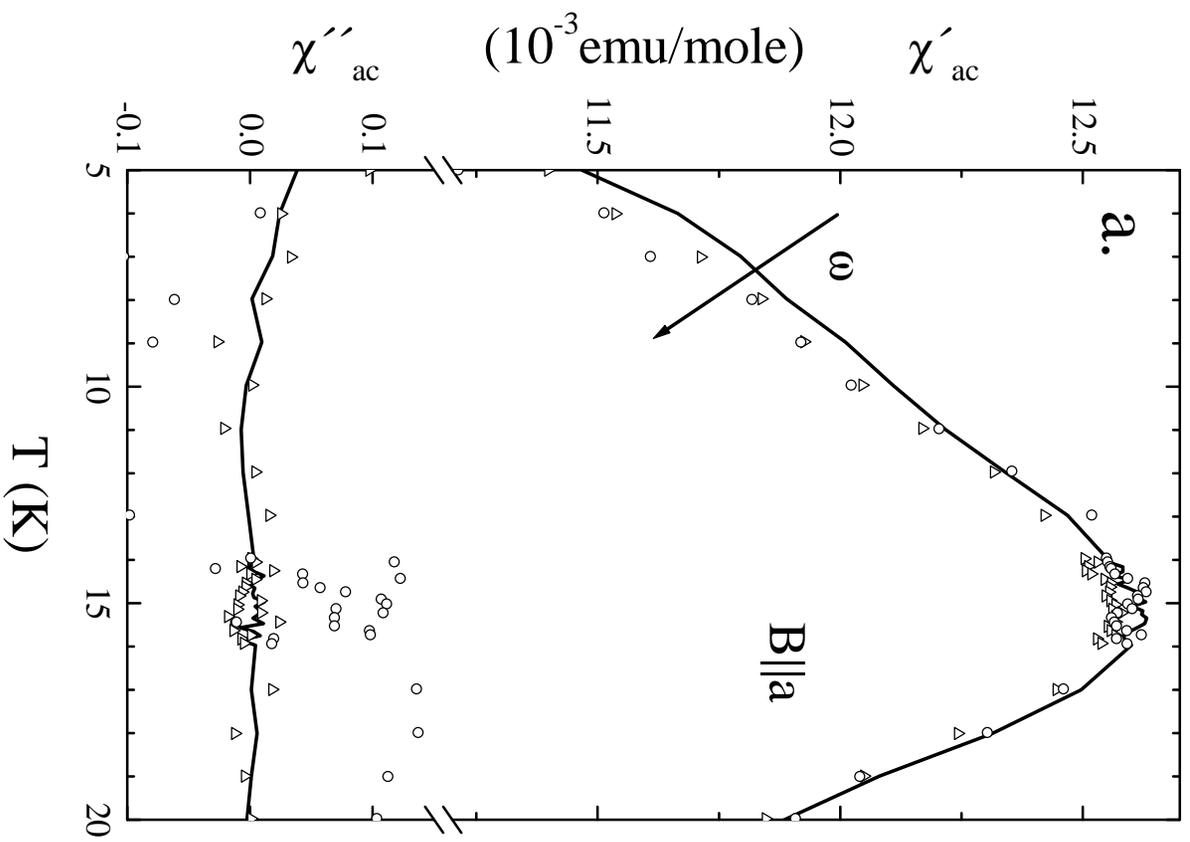
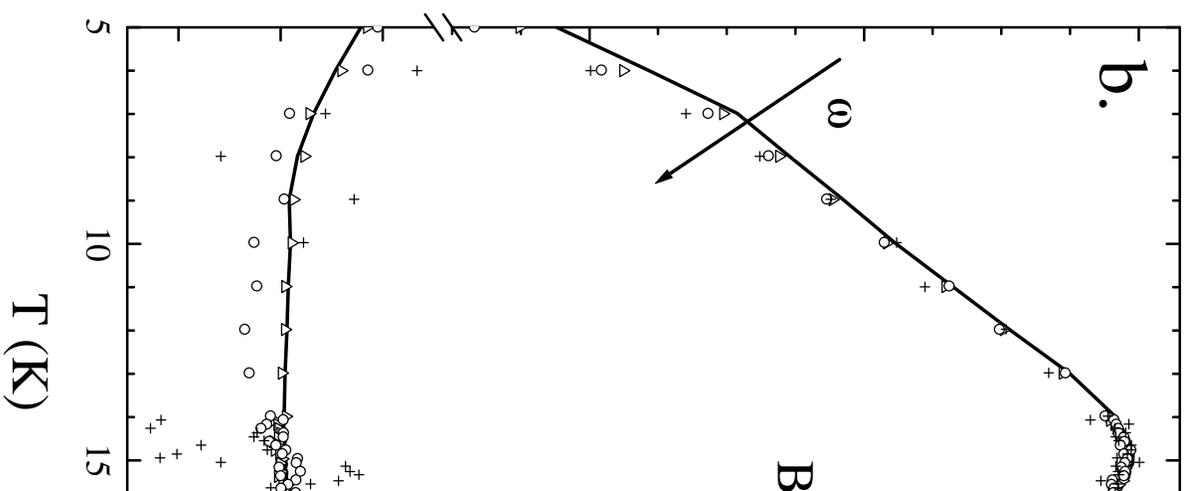
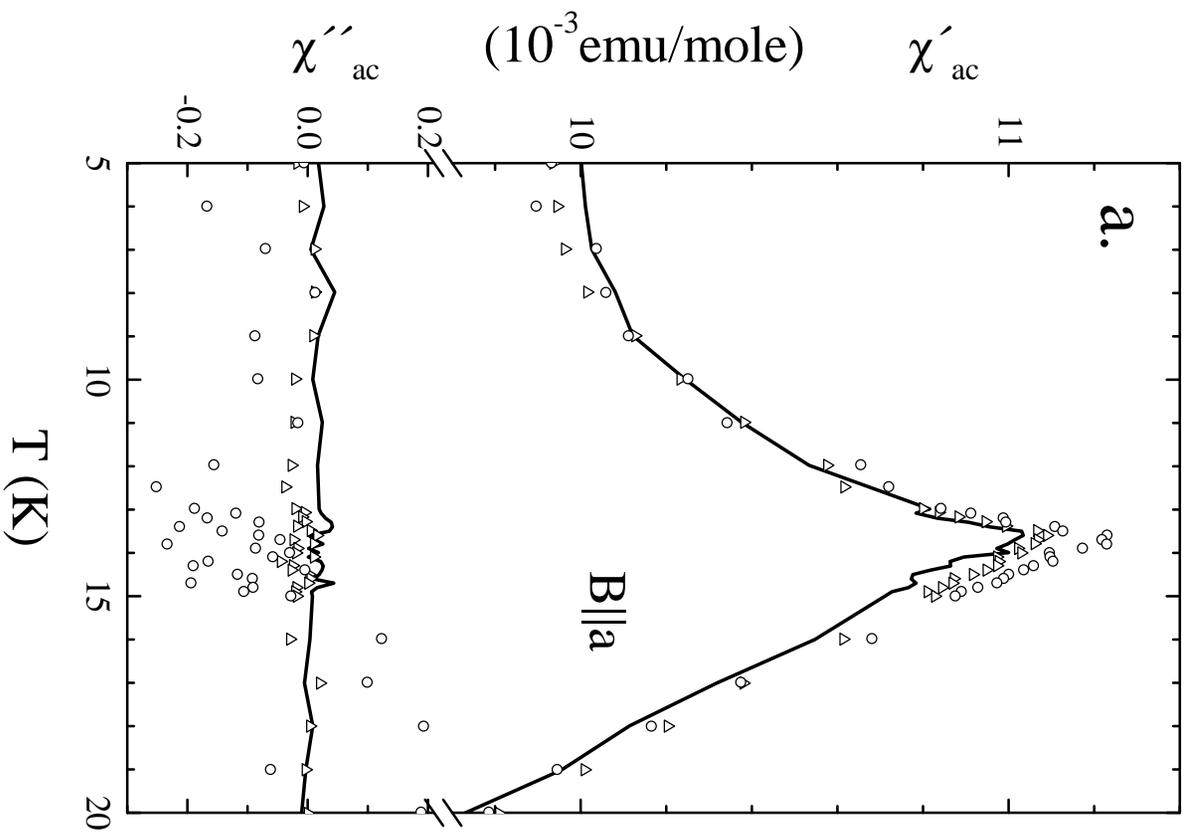
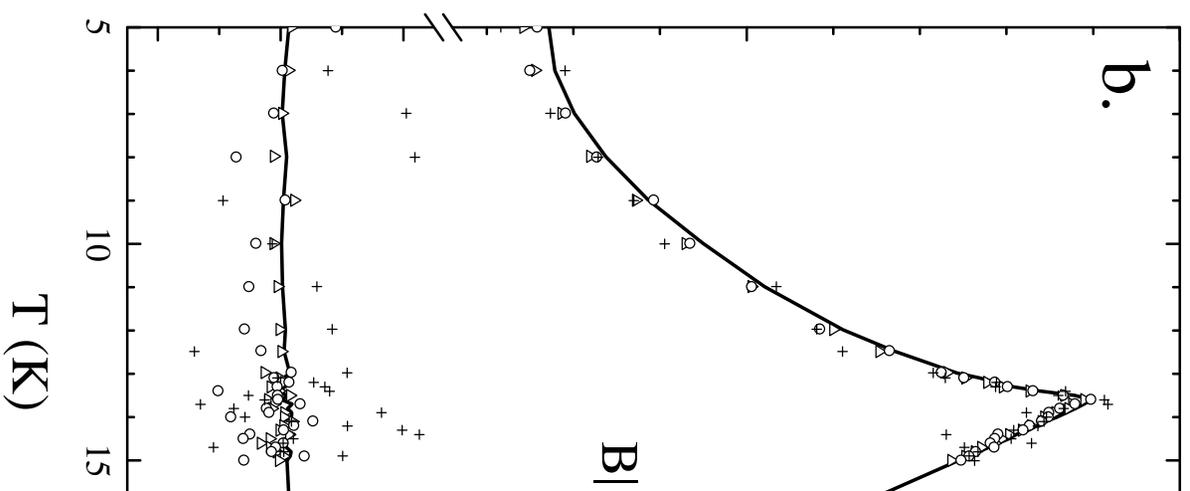
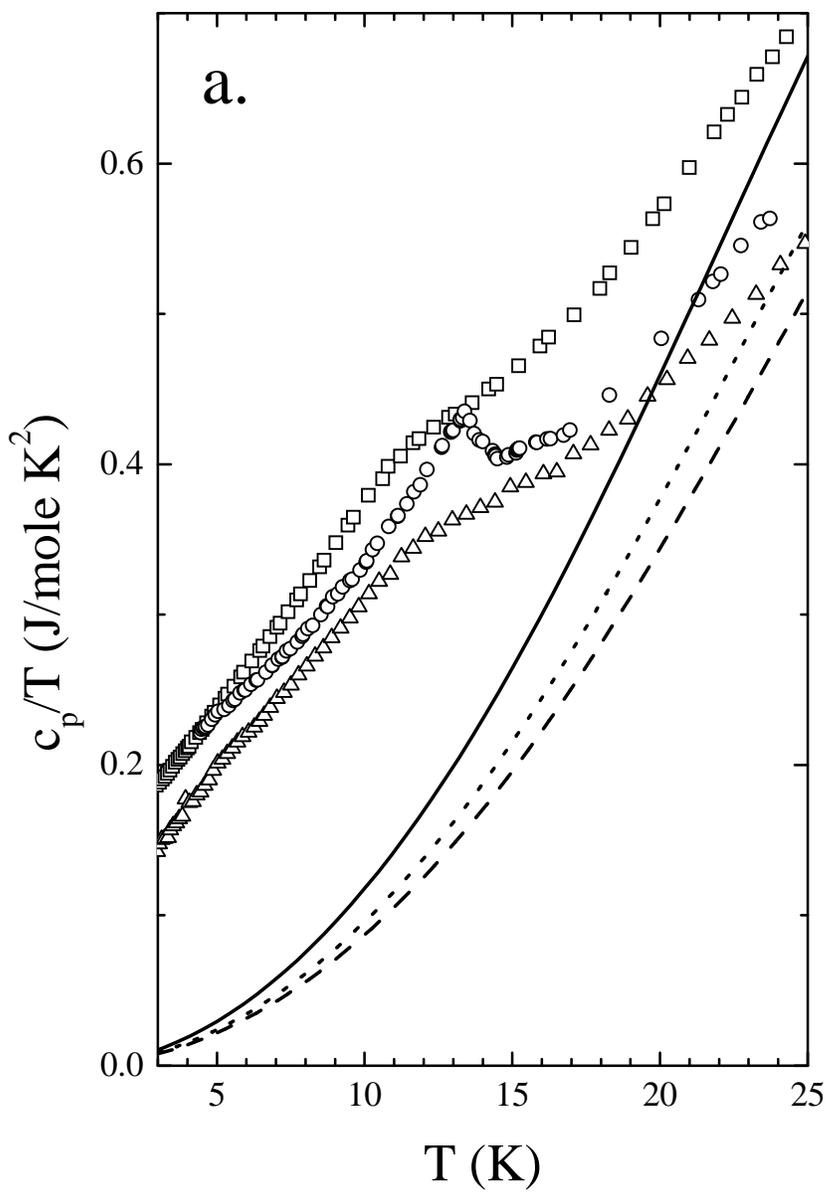
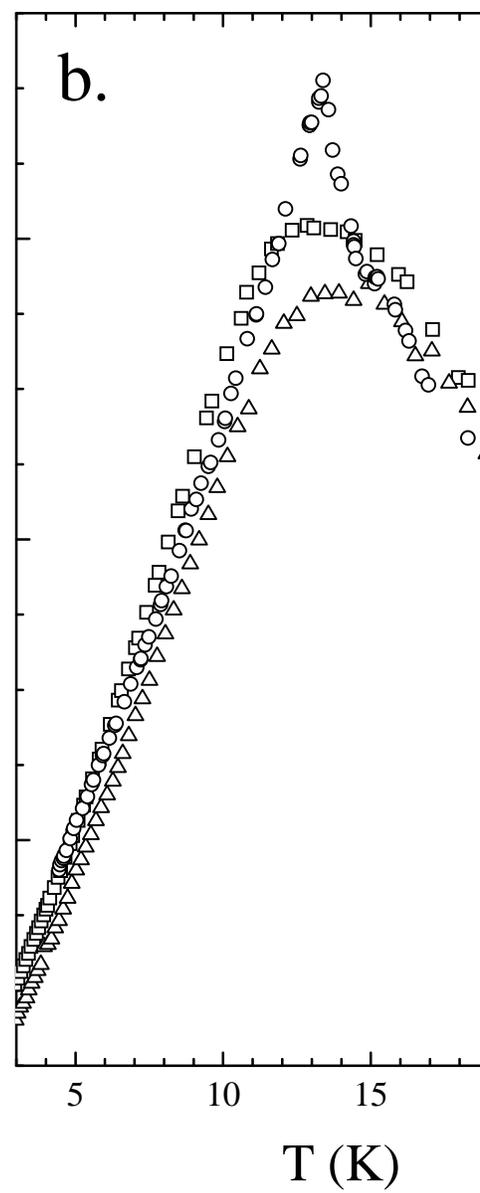
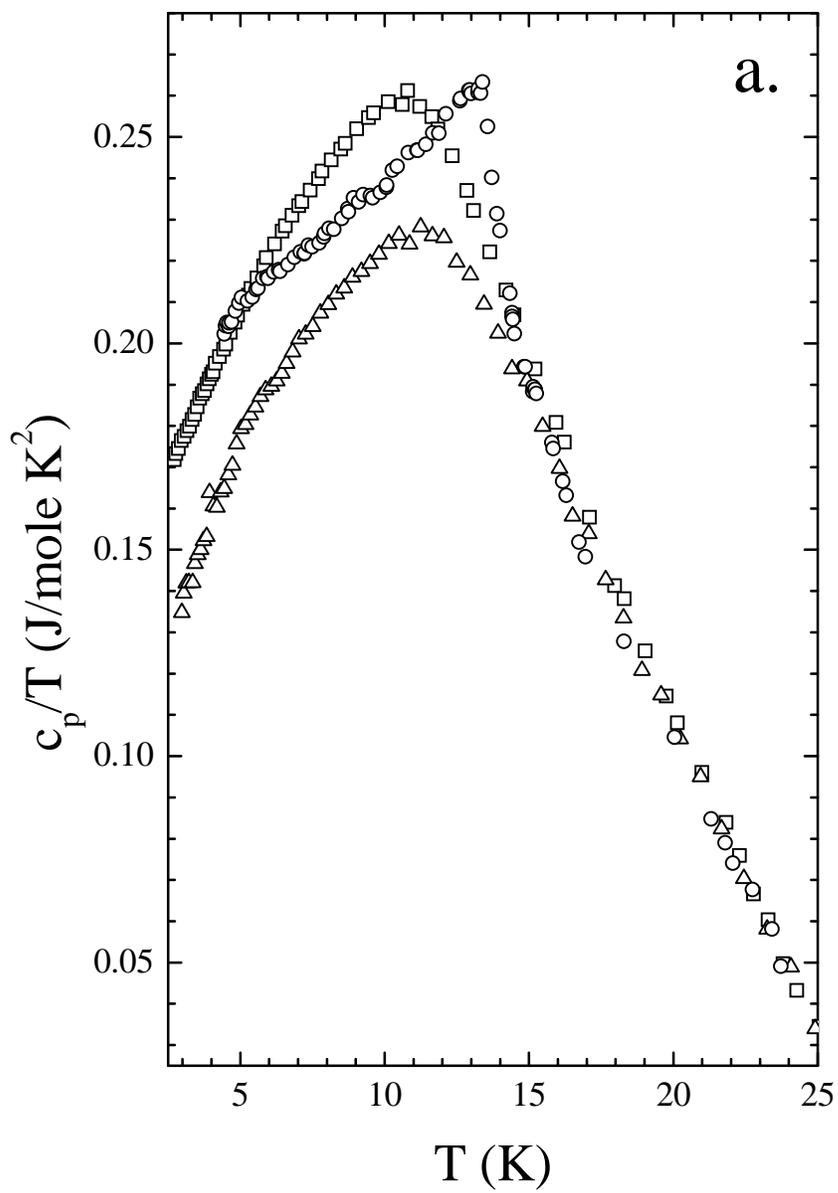
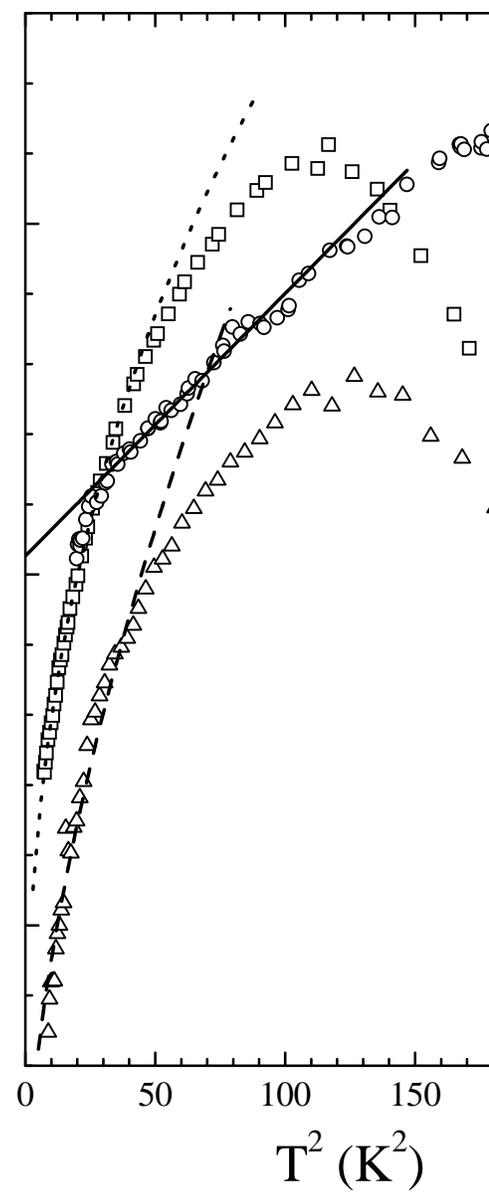
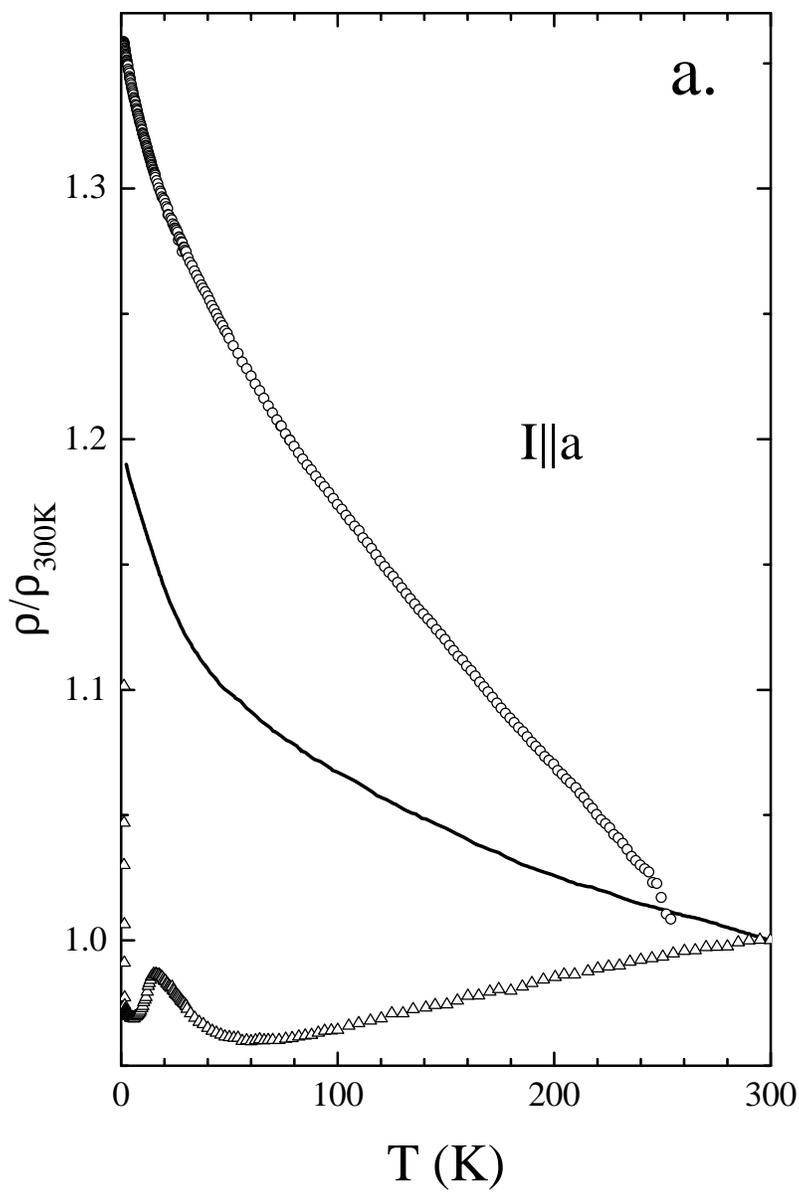
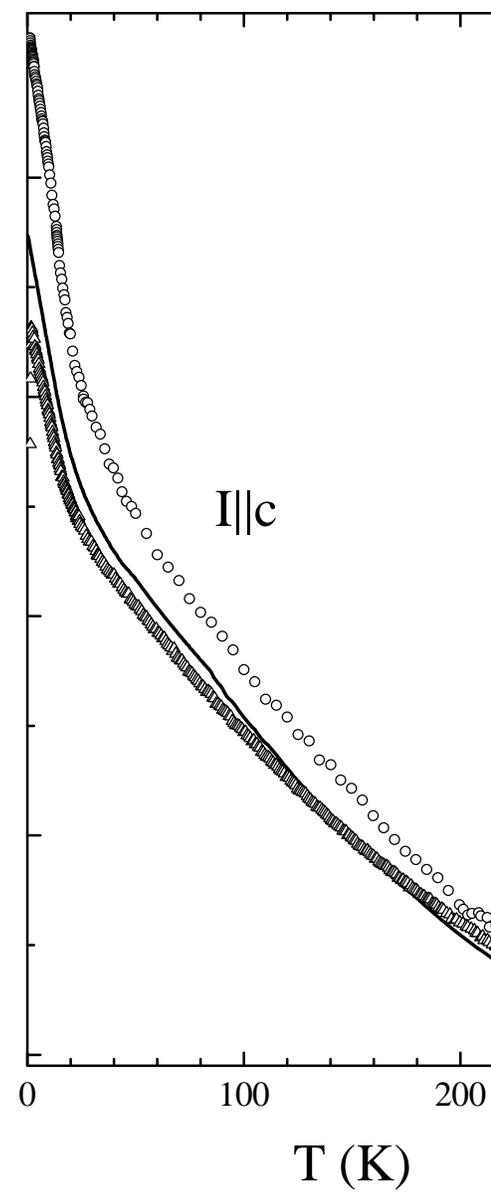
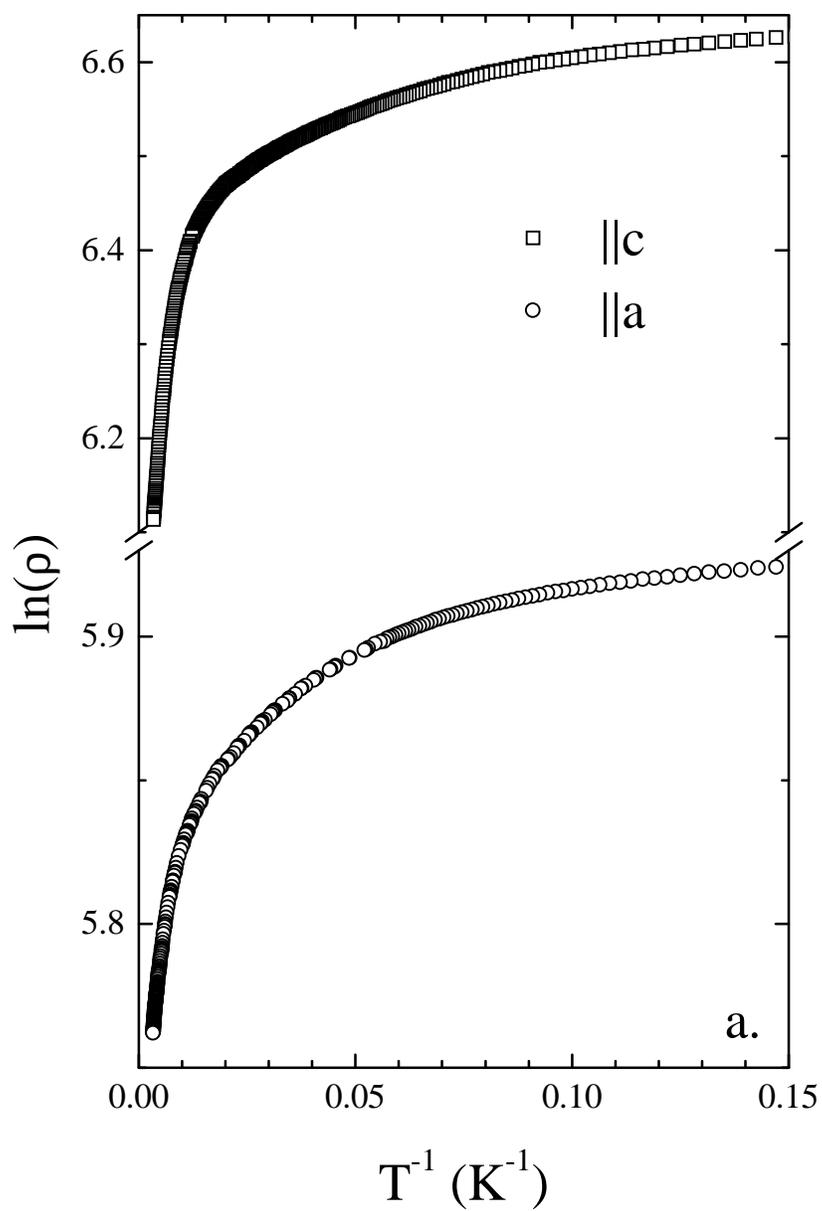
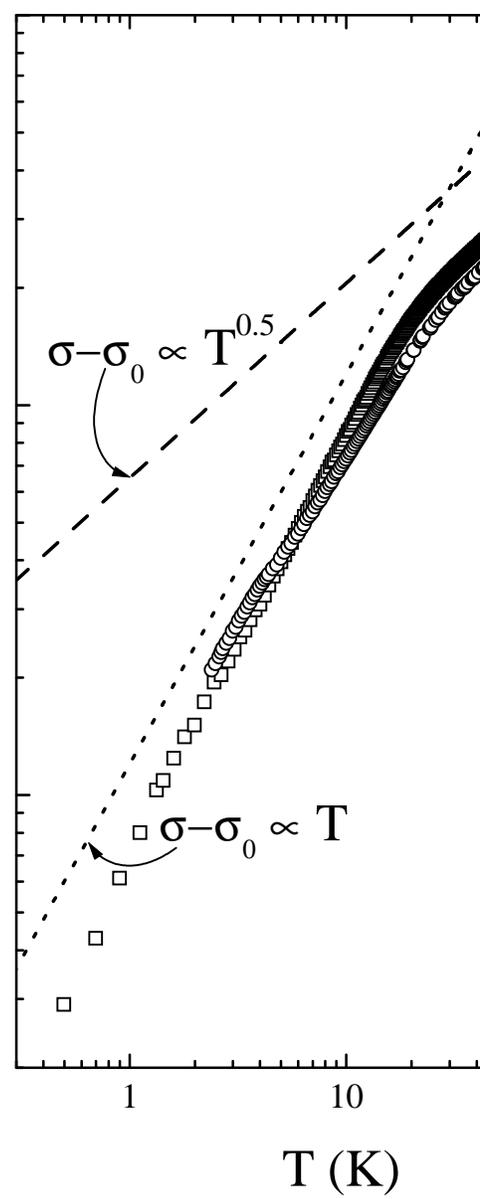
\end{document}